\documentstyle[spie]{article} 
\input{psfig.sty}

\title{Coupling of Length Scales and \\ Atomistic Simulation of MEMS Resonators}

\author{R. E. Rudd\supit{a} and J. Q. Broughton\supit{b} 
\skiplinehalf 
\supit{a} Department of Materials, University of Oxford, Oxford OX1 3PH UK \\
and SFA, Inc., 1401 McCormick Drive, Largo MD 20774 USA
\skiplinehalf 
\supit{b} Complex Systems Theory Branch, Naval Research Lab, Washington DC 20375-5345 USA \\
and Yale School of Management, New Haven CT 06520 USA
}

\authorinfo{Further author information: (Send correspondence to R.E.R.)\\R.E.R.:
 E-mail: robert.rudd@materials.oxford.ac.uk\\ J.Q.B.: E-mail: broughto@dave.nrl.navy.mil}

\begin{document} 
\maketitle 

\begin{abstract}

We present simulations of the dynamic and temperature dependent 
behavior of Micro-Electro-Mechanical Systems (MEMS) by utilizing
recently developed parallel codes which enable a coupling of length 
scales.  The novel techniques used in this simulation accurately
model the behavior of the mechanical components of MEMS down to 
the atomic scale.  We study the vibrational behavior of one class 
of MEMS devices:  micron-scale resonators made of silicon and quartz.  
The algorithmic and computational avenue applied here represents
a significant departure from the usual finite element approach based 
on continuum elastic theory.  The approach is to use an atomistic 
simulation in regions of significantly anharmonic forces and large 
surface area to volume ratios or where internal friction due to 
defects is anticipated.  This corrects the expected, but previously 
unquantified, failure of continuum elastic theory in the smallest 
MEMS structures.  Peripheral regions of MEMS which are well-described
by continuum elastic theory are simulated using finite elements for
efficiency.  Thus, in central regions of the device, the motion of 
millions of individual atoms is simulated, while 
the relatively large peripheral regions are modeled with finite elements. 
The two techniques run 
concurrently and mesh seamlessly, passing information back and forth.  
This coupling of length scales gives a natural domain decomposition, 
so that the code runs on multiprocessor workstations and supercomputers.
We present novel simulations of the vibrational behavior of micron-scale 
silicon and quartz oscillators.  Our results are contrasted with 
the predictions of continuum elastic theory as a function of size, and 
the failure of the continuum techniques is clear in the limit of small 
sizes.  We also extract the Q value for the resonators and study the
corresponding dissipative processes.

\end{abstract}

\keywords{MEMS, silicon resonators, coupling of length scales, atomistic simulation, molecular dynamics}

\section{The Failures of Standard Finite Elements for Small Devices}
\label{sect:intro}

The design of MEMS relies on a thorough understanding of the 
mechanics of the device itself.  
As system sizes shrink, MEMS are 
forced to operate in a regime where the assumptions of 
continuum mechanics are violated, and the usual finite element 
(FE) models fail.\cite{BMVK}  The behavior of materials begins 
to be atomistic 
rather than continuous, giving rise to anomalous and often 
non-linear effects:  
\begin{itemize}
\item The devices become less stiff and more compliant than FE predicts.
\item The roles of surfaces and defects become more pronounced.
\item Anharmonic effects become more important.
\item New mechanisms for dissipation become evident.
\item Statistical Mechanics becomes a key issue, 
even to the point that thermal fluctuations cannot be neglected.
\end{itemize}
The inadequacy of FE for these phenomena will be an obstacle 
to further miniaturization of MEMS.\cite{MSM}

\subsection{Micro-Resonator}

\begin{figure}[h]
\begin{center}
\begin{tabular}{c}
\psfig{figure=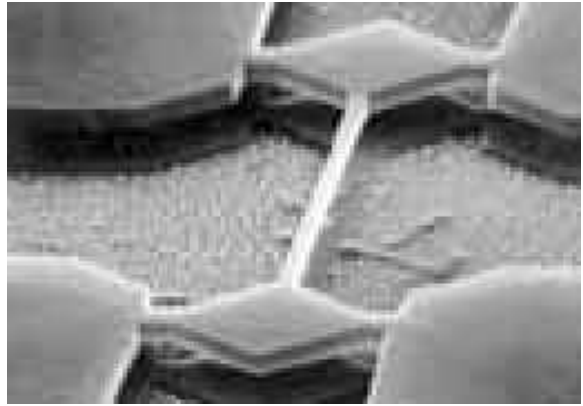,height=.3\textwidth} 
\end{tabular}
\end{center}
\caption[res] 
{ \label{fig:res}          
Silicon micro-resonator.  Length: approx.\ 0.2 microns.
Courtesy Prof.\ M.\ Roukes, Cal Tech. \cite{Roukes} }

\end{figure}

The failings of continuum elastic theory are evident in many ways 
for micro-resonators. 
Gigahertz resonators (cf. Fig.\ \ref{fig:res}) are roughly of the size 
$0.2 \times 0.02 \times 0.01$ microns.\cite{Roukes}
Devices of this size and smaller 
are so miniscule that materials defects and 
surface effects can have a large impact on their performance.   
Atomistic surface processes which would be negligible in large
devices are a major source of dissipation in sub-micron devices,
leading to a degradation in the Q-value of the resonator.
Additionally, bond breaking at defects can produce plastic deformations.  
These effects vary with temperature, and in the smallest devices,
the atomicity shows up through stochastic noise.
Systems smaller than about 0.01 microns are too small to be in 
the thermodynamic limit, and 
anomalous statistical mechanical effects are important.  These
effects are beyond continuum elastic theory.

\subsection{Micro-Gears}

\begin{figure}[h]
\begin{center}
\begin{tabular}{c}
\psfig{figure=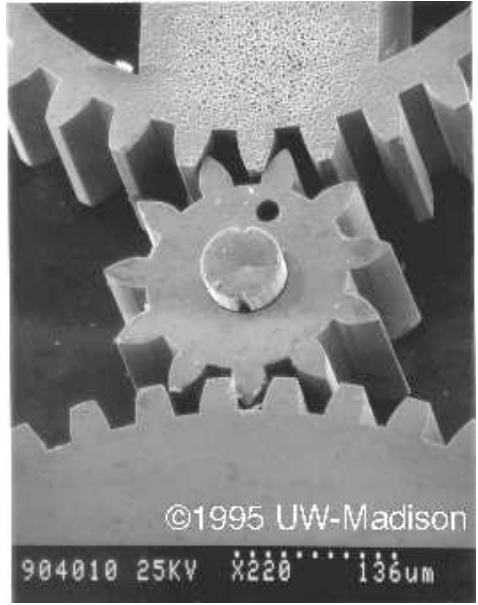,height=8cm} 
\end{tabular}
\end{center}
\caption[gear] 
{ \label{fig:gear}          
Example of a MEMS nickel gear train.  Dimensions: approx.\ 200 microns.
Courtesy Prof.\ H.\ Guckel, University of Wisconsin. \cite{Guckel} } 

\end{figure}

The anomalies are also evident in articulated devices.
The effects of wear, lubrication and friction can be expected to have
profound consequences on the performance of micron-sized machines,  
where areas of contact are a significant part of the 
system.  An archetypical example is the gear train,\cite{Guckel} something
at the heart of many micro-machines of the future (See Fig.\ \ref{fig:gear}).  
The process
of micro-gear teeth grinding against each other cannot be simulated
accurately with FE.  All of the failings listed above are evident.
The teeth are predicted to be too rigid.  Large amplitude high-frequency 
resonant modes may be missed.  Bond breaking and formation at the point
of contact can only be treated empirically. 

\section{Coupling of Length Scales}

Sub-micron MEMS require atomistic modeling, but atomistic simulations
are very computer intensive.  The
active region of a device whose dimensions approach a micron may
be modeled with atomistics, but
it is beyond the capabilities of even the largest supercomputers
to model the entire region of interest of a micron-scale device.
The coupling of the active region to the substrate is important,
and this may involve volumes of many cubic microns and billions
of atoms.  Thus, many sub-micron MEMS are too small for
finite elements and too large for atomistics.

We have developed a novel hybrid methodology which solves this 
problem.\cite{colsMeth}
Surfaces and other regions of micro-gears and micro-resonators 
in which atomistic
effects are critical can be modeled accurately with an atomistic 
simulation.  Finite elements offers an adequate and efficient 
model of other regions such as the body and the axle of the micro-gears
and the peripheral regions of the micro-resonator.  Then the two,
finite elements and atomistics, are melded together to run
concurrently through consistent boundary conditions at the 
handshaking interface.  Our multiscale
algorithm combines atomistic, finite element and 
even electronic simulations into a seamless, self-consistent 
monolithic simulation.\cite{colsMeth}  This {\em coupling of length scales} 
strikes a balance between computational accuracy and efficiency.  
This is part of
our DOD HPC Grand Challenge project to model
the dynamical behavior of MEMS.  The project gives us vast
computational resources at the Maui Supercomputer Center, where
we are developing codes to simulate the behavior of the next
generation of MEMS.  

\subsection{Hybrid of electronic structure, atomistics and continuum}

Our codes employ electronic and atomistic
simulation in the regions of bond-breaking and bond-deformation (large strain), 
so they are free from the assumptions of continuity
that lead to the failure of FE.  Instead, the codes are based
on well-tested interatomic potentials and electronic 
parameterizations.  The simulation tracks the motion of
each individual atom as it vibrates or perhaps 
diffuses in thermal equilibrium.  This means that the simulation 
automatically includes a wide range of phenomena.
The increased compliance at small sizes arises naturally, 
and when the stresses are large enough to induce
plastic deformation, this is simulated, too. 
No special phenomenology is required for 
effects due to surface relaxation, bond breaking and asperities.  

The technique involves a state-of-the-art
atomistic simulation (molecular dynamics, MD)
augmented self-consistently with concurrent
FE and electronic simulations (tight-binding, TB).\cite{colsMeth}  
This coupling of length scales 
is a novel finite temperature technique in materials simulation.  It
has never been attempted before, especially on parallel machines, 
and it
allows the extension of an essentially atomistic simulation to much 
larger systems.  Effects such as bond-breaking, defects, internal strain,
surface relaxation, statistical mechanical noise, and dissipation
due to internal friction are included.  The trick is that the gear train
is decomposed into different regions, FE, MD and TB, according to
the scale of the physics within that region.  

Our current codes will 
not run on existing desktop workstations, although they may in
five years.  Several factors will make this possible.  First, the
intrinsic speed and capacity of workstations will increase
exponentially with time (Moore's Law).  Second,  multiprocessor
workstations will become more common, so parallel codes will
offer the same advantage to workstations that they currently
offer only to supercomputers.  And third, our codes will become
more efficient as we continue to make algorithmic advances.

The ultimate goal of this work is to understand the important
atomistic effects both qualitatively and quantitatively.  To 
quantify the effects, we have calculated new constitutive 
relations appropriate for sub-micron resonators.  They include 
terms that describe atomistic surface effects, and other 
contributions to the energy of the structure.  These constitutive 
relations provide the starting point for continuum and finite 
element models.  Of course, the resulting models do not contain
all of the information from the atomistic simulation, so they
are only valid for a restricted range of device sizes and 
geometries, but the resulting finite element computations
could be carried out on a workstation.  This may be very
useful for device design.


\begin{figure}[h]
\begin{center}
\begin{tabular}{c}
\psfig{figure=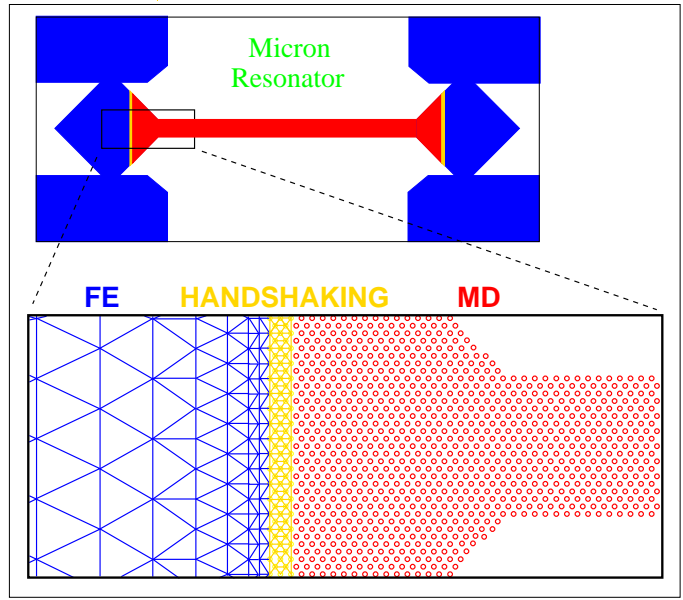,height=8cm} 
\end{tabular}
\end{center}
\caption[cols] 
{ \label{fig:cols}          
Schematic diagram of Coupling of Length Scales for a micro-resonator.
An atomistic simulation (MD) is used in the regions of the device
with moderate strain oscillations, while finite elements
(FE) is used in the peripheral regions where changes in the
strain are small.  The two are joined through a consistent
boundary condition in the handshaking region, and both are
run concurrently in lock-step.}

\end{figure} 

\subsection{Coarse-grained molecular dynamics}

The coupling between atomistics and finite elements described above 
works very well for many applications.  For example, in simulations
of crack propagation in silicon, the strain fields and elastic waves
emanating from
the crack tip pass fairly smoothly from the atomistic region into the 
finite element region.\cite{silCrack}  In particular, there is little
coherent backscatter of elastic shock waves from the MD/FE interface,
a problem that has plagued pure atomistic simulations of cracks.\cite{Holian}
It is clear, however, that finite
elements does not connect perfectly smoothly with atomistics in
the limit that the element size becomes atomic scale.  Finite element
analysis 
assumes that the energy density is spread smoothly throughout each
element, but at the atomic scale, the potential energy is localized
to the covalent bonds of silicon and the kinetic energy is localized
largely to the nuclei.  This small atomic scale mismatch can cause
problems in some types of simulation.

We have developed a substitute for finite elements which does connect
seamlessly to molecular dynamics in the atomic limit. It also reproduces
the results of finite elements (with slight improvements) in the
limit of large element size.  This new methodology is called
Coarse-Grained Molecular Dynamics (CGMD).\cite{CGMD}

CGMD has been constructed to provide a consistent treatment of 
the short wavelength modes which are present in the underlying
atomistics but are missing from the coarse finite element mesh.
These modes can participate in the dynamics and the thermodynamics
of the device.  In many situations, the short wavelength part
of the spectrum is relatively unpopulated, and the missing modes
are irrelevant to the behavior of the device.  But this is not
true for sufficiently small devices, or when there is a strong
source of high frequency elastic waves (such as a propagating
crack or when two micro-gears grind against each other).  In
addition to offering well-behaved thermodynamics, CGMD also
models the elastic wave spectrum more accurately than conventional
finite elements.  Furthermore, CGMD includes non-linear effects
that are compatible with the atomistics; i.e. it effectively
provides non-linear constitutive equations that are derived
from the atomistics without any free parameters.  

\section{Technical Approach}

\subsection{Micro-Resonator}

The goal of a multiscale simulation is to balance accuracy with
efficiency in an inhomogeneous system.
An atomistic simulation is employed in regions of significantly
anharmonic forces and large surface area to volume ratios or where 
internal friction due to defects is anticipated.  As shown in 
Fig.\ \ref{fig:cols}, the atomistic simulation models the central
region of the resonator.  This 
corrects the expected, but previously unquantified,
failure of continuum elastic theory in the smallest MEMS structures.
Regions of the micro-resonator which are well-described by continuum elastic
theory are simulated using finite elements.  These peripheral
regions include the coupling to the outside world, a substrate at
a given temperature.  Electronic structure calculations are used
in regions of very large strain and bond breaking, such as at
defects.  The electronic, atomistic and continuous regions
are joined seamlessly to form the complete simulation.

\begin{figure}[h]
\begin{center}
\begin{tabular}{c}
\psfig{figure=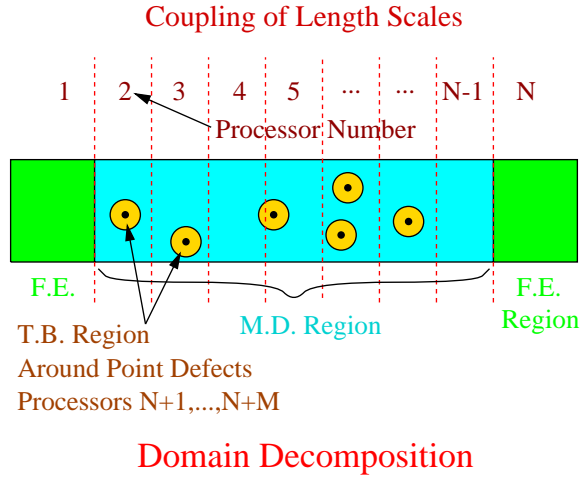,height=0.35\textwidth} 
\end{tabular}
\end{center}
\caption[resDomain] 
{ \label{fig:resDomain}          
Illustration of the domain decomposition of the long, thin
resonator showing coupling of length scales.
The smallest regions are electronic structure simulations
of the vicinity of defects, implemented with tight-binding (T.B.).
The intermediate regions
are molecular dynamics (M.D.) simulations and the end caps
depict part of the large finite element (F.E.) simulation.
The simulation is decomposed according to the scale of the
physics in each region, and then the simulation is distributed among
the processors on a supercomputer in order to optimize the overall
efficiency.}

\end{figure} 

Coupling length scales for the micro-resonator is accomplished as follows.
Any defects, the only regions of breaking bonds, are of necessity
described by electronic structure methodology. These are coupled to the
statistical mechanics
of sub-micron system sizes (to provide necessary fluctuations) via
conventional  molecular dynamics. This region, in turn, is coupled to 
micron and larger scales via finite elements.  
In each case the coupling between the regions amounts to a set
of consistent boundary conditions that enforce continuity.\cite{colsMeth}

This formulation of
the coupling of length scales gives a natural domain decomposition
to divide the computational load among parallel processors, as
shown in Fig.\ \ref{fig:resDomain}.  
The MD region is partitioned lengthwise into subregions, each of which is
assigned to a separate processor.
The FE regions (one at each end of the resonator) at present 
comprise one processor 
each, since they are not excessively computationally intensive. 
Each electronic structure domain consisting of about 20 atoms
is assigned to a separate processor.
The MD region uses an empirical potential
suitable for the material of interest; in this case, since the materials
are silicon and quartz, the potentials are those due to 
Stillinger and Weber \cite{SW} and Vashishta \cite{Nakano}, respectively.

Our simulations have been limited to relatively small defect densities,
so the multi-million atom atomistic region of the resonator requires
the majority of the processors.  Even though the electronic structure
calculation is intrinsically a much more expensive computation, the
total TB expense is less because there are relatively few TB atoms. 
Our codes are written in FORTRAN and MPI and run on IBM SP2s, allowing
atomistic regions of millions of atoms coupled, of course, to
large finite element regions, if necessary.

\subsection{Micro-Gears}

Fig.\ \ref{fig:gear} shows an example 
of micro-gear technology. \cite{Guckel} Such devices
can presently be made on the 100 micron scale and rotate at speeds 
of 150,000 RPM. Materials may be either polysilicon or nickel, depending 
upon the method of manufacture; we concentrate on silicon. We can 
expect next-generation devices to reach the 1 micron level. The speed 
with which they could be made to rotate is a subject for our research.

\begin{figure}[h]
\begin{center}
\begin{tabular}{c}
\psfig{figure=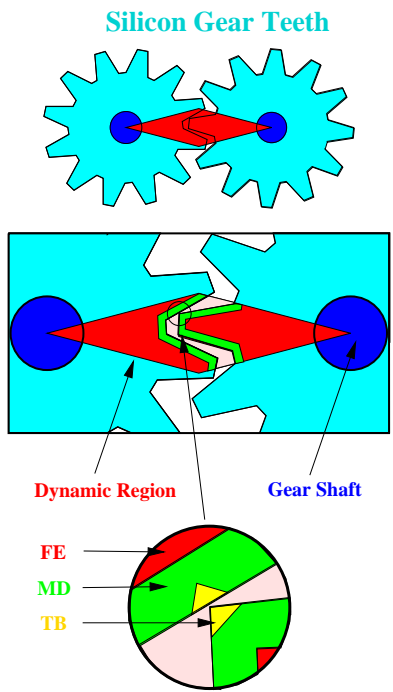,height=9cm} 
\end{tabular}
\end{center}
\caption[gearDomain] 
{ \label{fig:gearDomain}          
Illustration of dynamic simulation zone and
domain decomposition for coupling of length
scales:  from continuum (FE), to atomistics (MD) to electronic structure (TB).}

\end{figure} 

Fig.\ \ref{fig:gearDomain} illustrates the multiscale decomposition 
for the micro-gear train.
An inner region including the shaft 
is treated by finite elements.
FE uses the energy density that comes 
from constitutive relations
for the material of interest to produce a force on each nodal point
which drives the displacement field at that point using an algorithm
which looks just like molecular dynamics. 
We are also developing improved codes based on 
CGMD.\cite{CGMD}
The timestep used by the FE region has to be in
lock-step with the MD region (and also therefore with the TB region).
The handshaking between the FE and MD region is accomplished using 
a self-consistent overlap region.

Lastly, in regions at the gear-gear contact point in the non-lubricated
case, a tight-binding description is used.
When we study the lubricating properties of SAMs, the TB region 
is not required, since bond breaking is not an issue. 
Each TB region spreads across multiple processors.
Tight binding is a fast electronic structure method---with 
careful parameterization, it can be very accurate. We use the 
parameterization due to Bernstein and Kaxiras \cite{EK}. 
The effectiveness
of this coupling of length scales has been demonstrated previously 
in simulations of a crack opening in silicon. \cite{silCrack} 


\begin{figure}[h]
\begin{center}
\begin{tabular}{c}
\psfig{figure=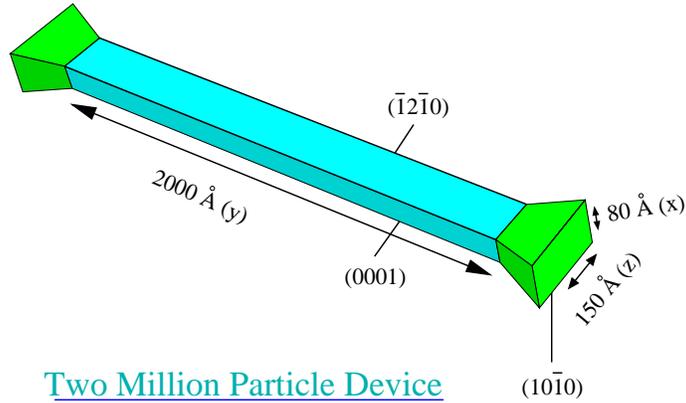,height=0.30\textwidth} 
\end{tabular}
\end{center}
\caption[qco] 
{ \label{fig:qco}          
Size and aspect ratio of the active region of a quartz oscillator, 
as simulated.}
\end{figure} 

\section{Results}

We have simulated micro-resonators with various sizes, defect concentrations
and temperatures, for comparison.  The dimensions of the largest oscillator
are shown in Fig.\ \ref{fig:qco}.  All of the devices have the same
aspect ratio, 25:2:1.  The motion of the resonator is simulated as
follows.  The initial configuration of the atoms is taken to be a single 
crystal of stoichiometric quartz in the desired device geometry 
(with some fraction of the atoms removed at random, 
if vacancies are to be modeled).  The system is brought to thermal
equilibrium in 100,000 time steps.  Then the resonator is deflected
into its fundamental flexural mode of oscillation, and released.
Once released, the thermostat is turned off, and no further energy
is put into the system.  

\begin{figure}[h]
\begin{center}
\begin{tabular}{c}
\psfig{figure=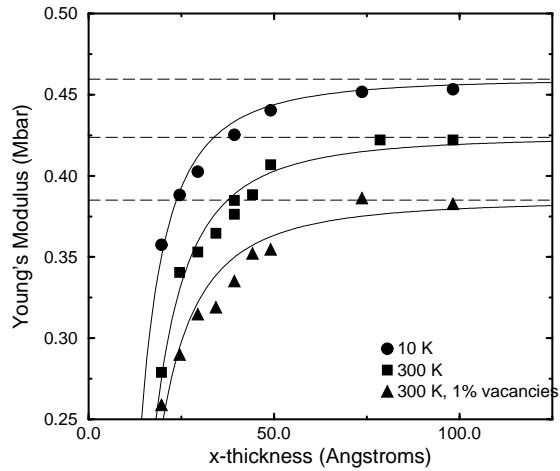,height=0.40\textwidth} 
\end{tabular}
\end{center}
\caption[young] 
{ \label{fig:young}          
A plot of the Young's modulus as a function of the device size
for a perfect crystal at two temperatures, T=10K and T=300K,
and a crystal with 1\% vacancies at T=300K.  The dashed lines
indicate the bulk value of the Young's modulus.  The solid
curved lines are the best fit to a constitutive relation
which includes terms for atomistic surface effects.}
\end{figure}

Various properties of the resonator have been studied.  
Fig.\ \ref{fig:young} shows the Young's modulus as a function of
size.  The resonant frequency of the oscillator is proportional
to the square root of the Young's modulus.  The dashed lines in
the figure indicate the bulk value, whereas the solid lines indicate
the best fit to a constitutive equation which includes a surface
term.  Atomistic effects are clearly evident for devices less
than 0.2 microns in length.  Note that this is true even for the
single crystal device at T=10K.  This device is essentially a
perfect crystal, disrupted only by the surfaces.  But it is
the surface relaxation which produces deviations from the bulk
behavior.

\begin{figure}[h]
\begin{center}
\begin{tabular}{c}
\psfig{figure=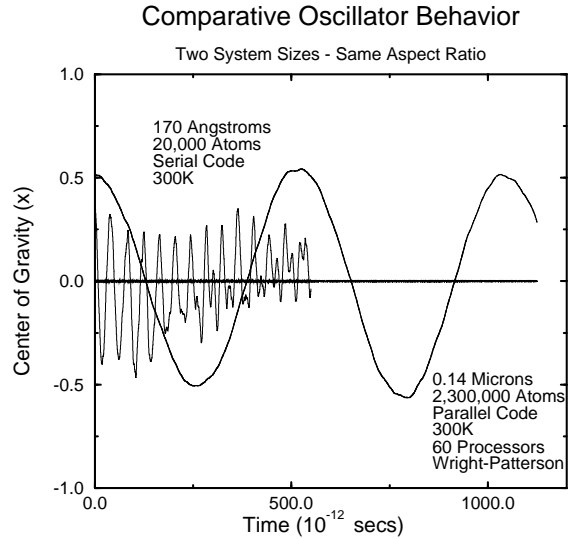,height=0.45\textwidth} 
\end{tabular}
\end{center}
\caption[QCO1] 
{ \label{fig:QCO1}          
A comparison of a 170\AA \ and a .14 micron 
quartz crystal oscillator.  The smaller system shows anharmonic and
surface effects.}

\end{figure}

\begin{figure}[h]
\begin{center}
\begin{tabular}{c}
\psfig{figure=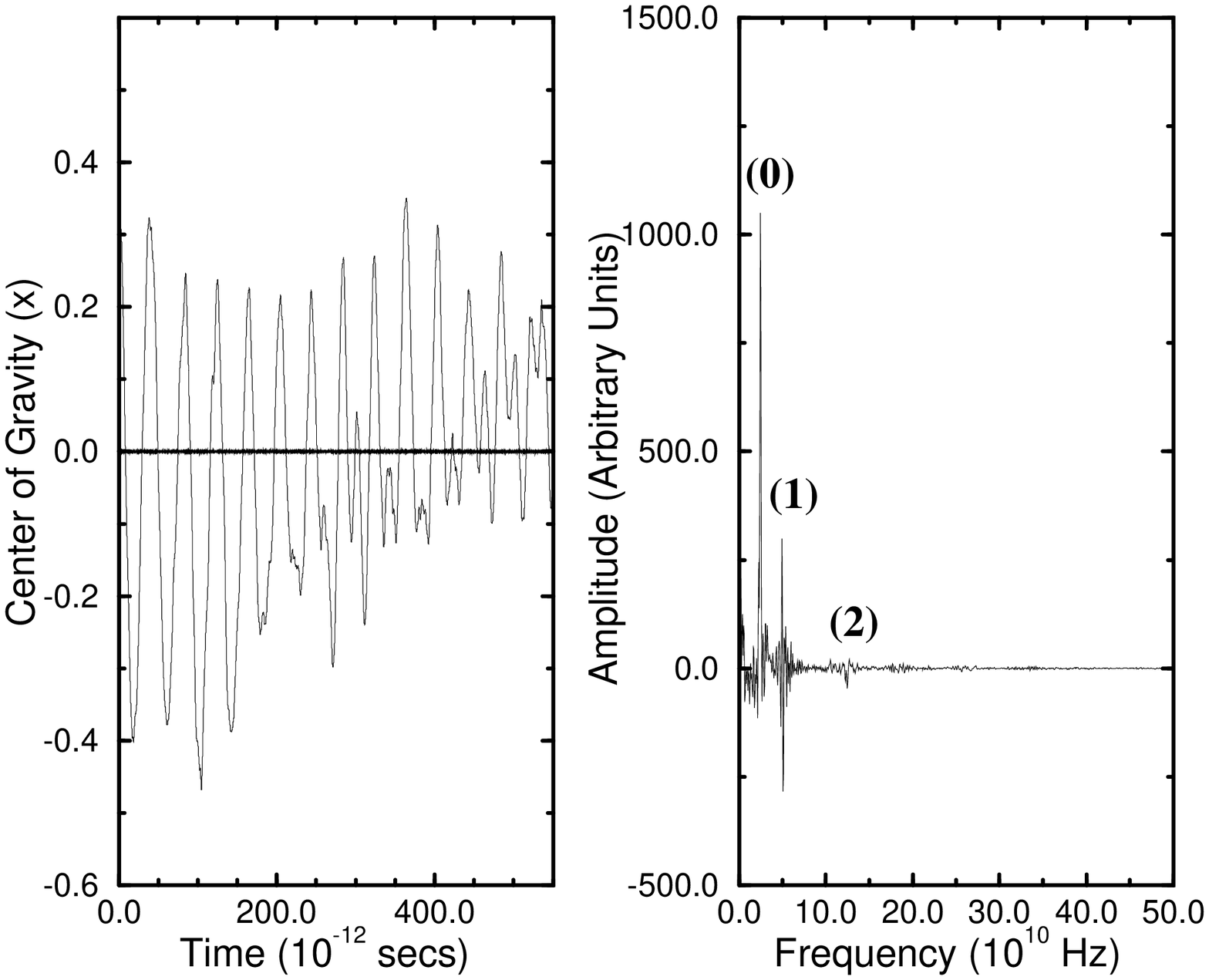,height=0.35\textwidth} 
\end{tabular}
\end{center}
\caption[ft] 
{ \label{fig:ft}          
A plot of the oscillations and corresponding frequency spectrum of the 0.017 
micron device at room temperature (T=300K).  The device is initially excited
in the fundamental mode (0).  Appreciable components
in the first (1) and second (2) harmonics have resulted from mode mixing
due to anharmonic lattice effects.}
\end{figure}

\begin{figure}[h]
\begin{center}
\begin{tabular}{c}
\psfig{figure=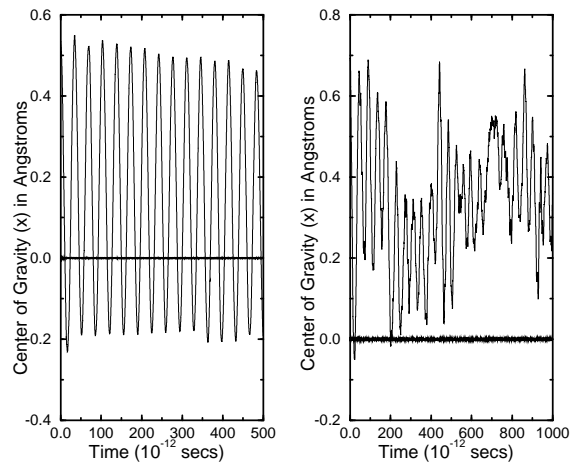,height=0.35\textwidth} 
\end{tabular}
\end{center}
\caption[QCO2] 
{ \label{fig:QCO2}          
Behavior of 1\% vacancy oscillator at two temperatures:
(a) T~=~10K and (b) T~=~300K.  Note that the oscillations
have shifted in the positive (x) direction in both cases,
due to a plastic deformation.}

\end{figure}

Fig. \ref{fig:QCO1} shows how
the oscillator rings as a function of time when plucked in flexural mode.
Note that relatively large deflections of the resonator are possible,
as great as 0.2\%, due to the increased compliance of the microscopic
devices.  The response of the oscillator at 300K shows marked effects
of anharmonicity \cite{BMVK}.  There is a pronounced frequency doubling
effect in the smaller oscillator, and even in the first few periods of
the larger oscillator there are clear departures from a sinusoidal
oscillation.  The mode mixing is most apparent in the Fourier transform
of the oscillations of the small device shown in Fig. \ref{fig:ft}.
A significant amount of the first harmonic (as well as a bit of the
second) has mixed into the spectrum.
This is not the case for an identical simulation run at T=10K,
where only the fundamental mode is present (not shown).\cite{BMVK}
Fig. \ref{fig:QCO2} shows that the response of the oscillator 
with 1\% vacancies is also
anomalous.  A substantial plastic deformation has resulted from the
presence of vacancies at both temperatures,
and again at 300K the response is highly anharmonic.

We can use simulations such as these to calculate the Q-value
for the various resonators.  The Q-value for the small devices at room
temperature shown in Figs. \ref{fig:ft} and \ref{fig:QCO2} are
Q=300 and Q=200, respectively, at their resonant frequency of
24 GHz.  
The simulation time for the
large system captures too few oscillations for a direct computation
of the Q-value, but a scaling analysis together with a fit to
the dominant dissipative processes enables us to estimate this
Q-value as well.  The results will be presented elsewhere.\cite{inProg}

This behavior could not be predicted from continuum elastic theory.  
The anharmonic response has been shown to be the result of surface effects.
The degradation of the Q-value of small resonators is also due to
a surface effect.  In large single crystal resonators, the primary
sources of dissipation are bulk thermoelastic and phonon-phonon 
processes.\cite{Braginsky}  Our results show that for small 
resonators at room temperature there is a new dominant source
of dissipation at the surface.  
The reduction of Q for small resonators has been observed
in experiments,\cite{Roukes,Kenny} where it has been attributed to
flawed surfaces.  Our results show that even for initially perfect
single crystal devices, atomistic effects cause a significant
degradation in Q at room temperature.

The plastic behavior is due to relaxation of the lattice about the
vacancies \cite{BMVK}.  The only way these effects could be addressed using
continuum elastic theory would be to construct empirical models that
would extend the standard finite element analysis.  However, this type
of model simply does not afford the confidence necessary to push the
frontiers of device design.  Our methodology, {\em coupling of length 
scales}, is able to make definitive predictions of device performance.

\section{Work in Progress}

We are in the process of setting up the micro-gear simulation.
We have chosen to study silicon gear
teeth because (a) a good TB parameterization exists and (b) because
gear teeth are often made of polysilicon.
The issues to be investigated are: (a) When two clean surfaces
are brought together they ``cold weld'' - thus when two teeth are in contact,
bonds will form across the opposing interfaces. As the gears rotate, these
bonds will break. How much matter is transferred? How rough is the newly
exposed surface? (b) Suppose we affix linear polymer chains to the surfaces
of the gears (self-assembled monolayers) of length (say) 12 carbon atoms,
does this significantly reduce wear and friction?
The monolayer molecules in question are the alkyltrichlorosilanes. Interatomic
potentials are available for similar polymeric systems \cite{IS}.
Technologically, such processing of the
surfaces is very doable \cite{SAM}, and in fact has already been applied to
micro-gears \cite{DCMS}. How much energy is dissipated into the body
of the gears when such surfaces rub?
At what speed can we run a gear train before
entanglement and relaxation times in the SAM polymer become an issue?



\section{Conclusion}

The algorithmic and computational avenue applied here represents
a significant departure from the usual finite element approach based 
on continuum elastic theory.  We have shown that atomistic simulation,
and in particular multiscale atomistic simulation, offers significant
improvements for the modeling of MEMS on sub-micron length scales.  
It is at these scales that 
some of the assumptions of continuum mechanics fail.  The issue is at 
what system size and in what way they fail. These are issues that 
we are able to answer unambiguously using atomistic simulation and
coupling of length scales. 

\acknowledgments     
 
This work was supported by ONR and DARPA. Supercomputing resources
were provided through a DOD HPC `Grand Challenge' award at the
Maui and ASC Supercomputing Centers.


\bibliography{dtm99Bib}   
\bibliographystyle{spiebib}   
      
\end{document}